\documentclass[10pt]{article}

\usepackage{amsmath}
\usepackage{amssymb}
\usepackage{graphicx}
\usepackage{psfrag}
\usepackage{color}
\usepackage{sistyle}
\usepackage{cite}

\renewcommand{\Im}{\mathrm{Im}}

\begin{document}

\title{Thermal noise of microcantilevers in viscous fluids}
\author{L. Bellon\\
Universit\'e de Lyon\\
Laboratoire de Physique, ENS Lyon, CNRS UMR5672\\
46 all\'ee d'Italie, 69364 Lyon Cedex 07, France\\
Ludovic.Bellon@ens-lyon.fr}

\date{\today}

\maketitle

%%\noindent \emph{The following article appeared in \emph{Journal Of Applied Physics, {\bf %%104}, 104906} and may be found at \linebreak %%\emph{http://link.aip.org/link/?JAP/104/104906}. Copyright 2008 American Institute of %%Physics. This article may be downloaded for personal use only. Any other use requires prior %%permission of the author and the American Institute of Physics.}

\begin{abstract}
We present a simple theoretical framework to describe the thermal noise of a microscopic mechanical beam in a viscous fluid: we use the Sader approach to describe the effect of the surrounding fluid (added mass and viscous drag), and the fluctuation dissipation theorem for each flexural modes of the system to derive a general expression for the power spectrum density of fluctuations. This prediction is compared with an experimental measurement on a commercial atomic force microscopy cantilever in a frequency range covering the two first resonances. A very good agreement is found on the whole spectrum, with no adjustable parameters but the thickness of the cantilever.
\end{abstract}

\section{Introduction}

Cantilevers of micrometer size are nowadays present in many applications, ranging from chemical and biological sensors\cite{Lavrik-2004} to scanning probe microscopy\cite{Meyer-2004}. In many cases, these tiny mechanical systems operate in a fluid environment (air or water for instance) which has a great influence on their dynamical behavior: the viscosity of the medium will broaden the structural resonances while the added mass due to the fluid moving along with the cantilever will shift their frequencies. A few theoretical models have been proposed to account for these effects \cite{Sader-1998,Paul-2004,Green-2005,Paul-2006,Dorignac-2006,VanEysden-2007,Salapaka-1997,Elmer-1997,Maali-2005,Basak-2006}, and validated experimentally \cite{Salapaka-1997,Elmer-1997,Maali-2005,Basak-2006,Chon-2000,Burnham-2003,Ghatkesar-2008}. Among the prediction of these approaches, the power spectrum of thermal noise induced fluctuations is of particular interest for its applications in atomic force microscopy (AFM) or micro-electromechanical systems (MEMS). We will focus here on the approach by Sader et al. \cite{Sader-1998,Green-2005,VanEysden-2007}, and derive a generic formula for the thermal noise using the fluctuation dissipation theorem for each mode, extending the work of Paul and Cross \cite{Paul-2004} in a simpler framework than Dorignac et al.\cite{Dorignac-2006}. This careful analysis is indeed incorrect in the original work of Sader et al., as noted by Paul and Cross \cite{Paul-2004}, Dorignac et al. \cite{Dorignac-2006} and Sader et al. themselves (see ref. 44 of \cite{Green-2005}). We will then compare the prediction of the model to a noise measurement on a commercial AFM cantilever in air: an excellent agreement is obtained for the full frequency behavior of the two first resonances, as well as at low frequencies and in the intermediate region between the two pics.

\section{Theory}

\begin{figure}[h]
\begin{center}
\includegraphics{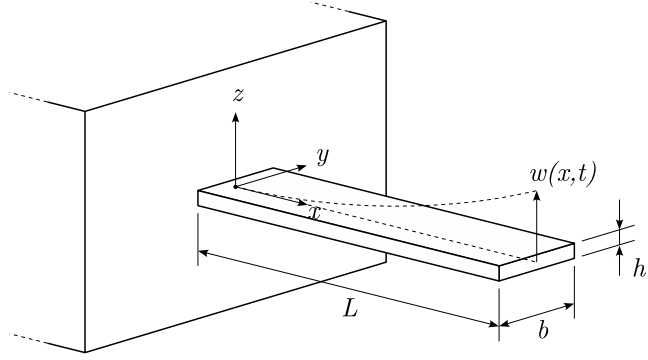}
\end{center}
\caption{Geometrical model for the cantilever: rectangular beam with length $L$, width $b$ and thickness $h$. Deformation are supposed to be only along axis $z$ and to depend only of time $t$ and spacial coordinate $x$, they are described by the deflection $w(x,t)$.}
\label{Fig:cantilever}
\end{figure}

We will first recall the general theory to describe the motion of a cantilever immersed in a viscous fluid and subject to an arbitrary external forcing, using the same notations and approach as J. Sader \cite{Sader-1998}, and then derive the expression for the thermal noise of the mechanical system. The cantilever is sketched on Fig.\ref{Fig:cantilever}. Its length $L$ is supposed to be much larger than its width $b$, which itself is much larger than its thickness $h$. We will limit ourself in this study to the flexural modes of the cantilever: the deformations are supposed to be only perpendicular to its length (along axis $z$ of Fig.\ref{Fig:cantilever}) and uniform across its width. Theses deformation can thus be described by the deflection $w(x,t)$, $x$ being the spacial coordinate along the beam normalized by $L$, and $t$ the time. The equation of describing the dynamics of $w$ is \cite{Graff-1975}:
\begin{equation} \label{eq:beam}
\frac{E I}{L^{4}} \frac{\partial^{4}w(x,t)}{\partial x^{4}}+\mu \frac{\partial^{2}w(x,t)}{\partial t^{2}}=F(x,t)
\end{equation}
where $E$ is the Young modulus, $I=b h^{3}/12$ the second moment of inertia of the cantilever, $\mu$ the mass per unit length and $F$ the external force per unit length. The boundary conditions for eq.~\ref{eq:beam} correspond to the classical clamped ($x=0$) and free end ($x=1$) conditions:
\begin{align} \label{eq:boundary}
w(x{=}0,t)& = 0& \left . \frac{\partial w(x,t)}{\partial x}\right |_{x=0}& = 0 \\
\left . \frac{\partial^{2}w(x,t)}{\partial x^{2}}\right |_{x=1}& = 0& \left . \frac{\partial^{3}w(x,t)}{\partial x^{3}}\right |_{x=1}& = 0
\end{align}

In the Fourier space, we can rewrite eq.~\ref{eq:beam} as
\begin{equation} \label{eq:beam Fourier}
\frac{k}{3} \frac{d^{4}w(x | \omega)}{d x^{4}}-m_{c} \omega^{2} w(x | \omega)=F(x | \omega) L
\end{equation}
with $\omega$ the pulsation in the Fourier space, $m_{c}=\mu L$ the mass and $k=3 E I / L^{3}$ the spring constant of the cantilever. Solving this equation in absence of external forces leads to the well known normal modes of oscillation of a clamped beam in vacuum with spatial profiles
\begin{equation} \label{eq:normal modes}
\phi_{n}(x)=(\cos C_{n}x -\cosh C_{n}x) + \frac{\cos C_{n}+\cosh C_{n}}{\sin C_{n} + \sinh C_{n}}(\sin C_{n}x - \sinh C_{n}x)
\end{equation}
where $C_{n}$ is the $n^{th}$ solution of equation
\begin{equation} \label{eq:Cn}
1+ \cos C_{n}\cosh C_{n} = 0
\end{equation}
which leads to $C_{1}=1.875$, $C_{2}=4.694$, \ldots , $C_{n}=(n-1/2)\pi$. The normals modes
$\phi_{n}(x)$ for $n=1,\ldots,\infty$ form an orthonormal basis of the functions of $x$ in $[0,1]$ \cite{Graff-1975}:
\begin{equation} \label{eq:phi_n orthonormal}
\int_{0}^{1} \phi_{n}(x)\phi_{m}(x)dx=\delta_{n,m}
\end{equation}
The pulsations of these resonant modes are given by the dispersion equations
\begin{equation} \label{eq:omega_n} 
m_{c} \omega_{vac,n}^2 = \frac{C_{n}^4}{3} k 
\end{equation}

When the cantilever is moving inside a fluid, the force acting on the cantilever can be decomposed into 2 components: the force corresponding to the hydrodynamic load  $F_{\mathrm{hydro}}(x | \omega)$ due to the motion of the beam in the fluid, and the actual external driving $F_{\mathrm{drive}}(x | \omega)$. Following Sader's approach  \cite{Sader-1998}, the hydrodynamic load can be approximated by
\begin{equation} \label{eq:F_hydro}
F_{\mathrm{hydro}}(x | \omega) = \frac{\pi}{4} \rho \omega^{2} b^{2} \Gamma(\omega) w(x | \omega)
\end{equation}
where $\Gamma(\omega)$ is the hydrodynamic function corresponding to the rectangular cantilever beam and $\rho$ the density of the fluid. The real part $\Gamma_{r}$ of $\Gamma$ corresponds to the added mass due the fluid moving along with the cantilever during its motion (normalized to $m_{f}=\pi \rho L b^{2} /4$, the mass of a cylinder of fluid of diameter $b$ and length $L$), and the imaginary part $\Gamma_{i}$ to the viscous drag. To solve the equation of motion of the cantilever in the fluid, let us decompose the deflection $w(x | \omega)$ and external driving $F_{\mathrm{drive}}(x | \omega)$ on the orthonormal basis of the normal modes $\phi_{n}$:
\begin{eqnarray}
w(x | \omega) & = & \sum_{n=1}^{\infty} \beta_{n}(\omega) \phi_{n}(x) \label{eq:w=sum beta_n phi_n} \\
F_{\mathrm{drive}}(x | \omega) & = & \frac{1}{L} \sum_{n=1}^{\infty} \eta_{n}(\omega) \phi_{n}(x)
\end{eqnarray}
We note here that this decomposition is also valid in the real time space, with amplitude $\beta_{n}(t)$ and external driving $\eta_{n}(t)$ of each mode simply being the inverse Fourier transform of $\beta_{n}(\omega)$ and $\eta_{n}(\omega)$. These 2 variables are coupled by the hamiltonian $H$ of the system. Indeed, let us compute the infinitesimal work $\delta W$ of $F_{\mathrm{drive}}$ when the deflection changes by $\delta w$:
\begin{eqnarray}
\delta W & = & \int_{0}^{1} L dx F_{\mathrm{drive}}(x,t) \delta w(x,t)  \\
 & = & \sum_{n=1}^{\infty} \eta_{n}(t) \delta \beta_{n}(t)
\end{eqnarray}
For a reversible transformation, we can write $dH=\delta W$, hence
\begin{equation} \label{eq:coupled variables}
\frac{\partial H}{\partial \beta_{n}} = \eta_{n}
\end{equation}
This last equation shows that the amplitude $\beta_{n}$ and external driving $\eta_{n}$ of each mode are coupled variables by the hamiltonian of the system. We can thus apply the Fluctuation Dissipation Theorem \cite{deGroot-1984}: the power spectrum density $S_{\beta_{n}}$ of fluctuation of amplitude for mode $n$ is given by
\begin{equation} \label{eq:S_beta_n-FDT}
S_{\beta_{n}}(\omega)=\frac{2 k_{B} T}{\pi \omega} \Im(\frac{\beta_{n}(\omega)}{\eta_{n}(\omega)})
\end{equation}
where $k_{B}$ is the Boltzmann constant, $T$ the temperature of the thermostat and $\Im(.)$ is the imaginary part of its argument.

We will now explicit the right part of eq.~\ref{eq:S_beta_n-FDT}. Using the orthonormalization of the base $\phi_{n}$, we can easily show that eq.~\ref{eq:beam Fourier} leads to the following equation for each mode:
\begin{equation} \label{eq:mode n}
\left(\frac{k}{3} C_{n}^{4} -\left(m_{c}+m_{f}\Gamma(\omega)\right) \omega^{2}\right) \beta_{n}(\omega) = \eta_{n}(\omega)
\end{equation}
The amplitude $\beta_{n}$ of each mode is thus governed by the equation of motion of an oscillator with stiffness $k C_{n}^{4}/3$, mass $m_{c}+m_{f}\Gamma_{r}(\omega)$ and damping coefficient  $m_{f}\omega\Gamma_{i}(\omega)$, forced by external driving $\eta_{n}$. If $\Gamma(\omega)$ is a slow varying function of the frequency, close to its resonance each mode behaves like an harmonic oscillator. Using the above expression, eq.~\ref{eq:S_beta_n-FDT} can be rewritten as
\begin{align} \label{eq:S_beta_n}
S_{\beta_{n}}(\omega) &= \frac{2 k_{B} T}{\pi \omega} \Im(\frac{1}{\frac{k}{3} C_{n}^{4} -\left(m_{c}+m_{f}\Gamma(\omega)\right) \omega^{2}}) \\
&= \frac{2 k_{B} T}{\pi} \frac{1}{m_{c}} \frac{\tau \Gamma_{i}(\omega)  \omega}{(\omega_{vac,n}^{2}-(1+\tau \Gamma_{r}(\omega)) \omega^{2})^{2} +\tau^{2}\Gamma_{i}^{2}(\omega)\omega^{4}}
\end{align}
with $\tau=m_{f}/m_{c}$. In general, dissipation couples the modes calculated from the non-dissipative equations, but within the approximation of position independent dissipative forces (eq~\ref{eq:F_hydro}), the noises are uncorrelated. We can thus write a generic expression of the power spectrum density of the deflection using eq.~\ref{eq:w=sum beta_n phi_n}:
\begin{equation} \label{eq:PSDw}
S_{w}(x,\omega) = \sum_{n=1}^{\infty} S_{\beta_{n}}(\omega) |\phi_{n}(x)|^{2}
\end{equation}
This last equation is in agreement with the one derived by Dorignac et al. \cite{Dorignac-2006} in a more complex framework.

\section{Experiment}

To test the validity of the above formalism, we measure the thermal noise of a commercial AFM cantilever (Budget Sensor BS-Cont) with the following nominal geometry: $L=\SI{450}{\micro m}$, $b=\SI{50}{\micro m}$, $h=\SI{2}{\micro m}$. The cantilever is made of Silicium and is coating-less, thus for theoretical expressions we used the tabulated values of Si for the young Modulus ($E=\SI{169}{GPa}$ for Si$_{110}$, the usual crystalline orientation along the length of AFM cantilever) and density ($\rho_{c}=\SI{2340}{kg.m^{-3}}$). The length and width were checked with an optical microscope, and the thickness is inferred from the resonance frequency of the cantilever (from eq.~\ref{eq:omega_n}), which eventually leads to $L=\SI{450}{\micro m}$, $b=\SI{48}{\micro m}$, $h=\SI{2.07}{\micro m}$.

\begin{figure}[h]
\begin{center}
\includegraphics{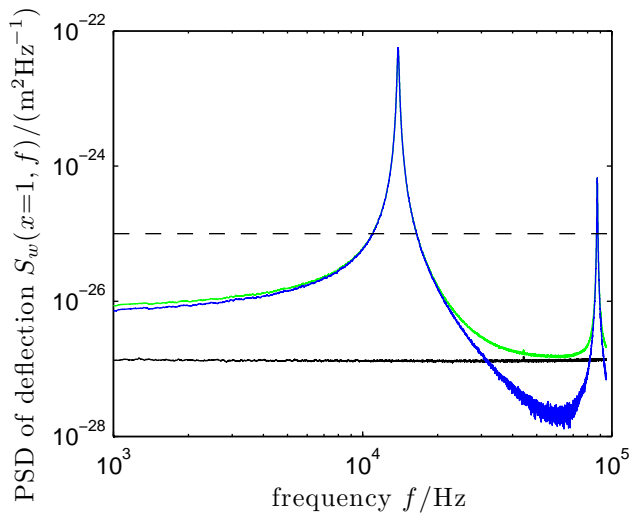}
\end{center}
\caption{Power spectrum density (PSD) of deflection. The PSD of thermal noise driven deflection at the free end of the cantilever \mbox{({\bf \textcolor{green}{--}})} is measured with an home made interferometric setup in a $\SI{1}{kHz}$ to $\SI{100}{kHz}$ range. We clearly see the two first flexural modes of the mechanical beam, well above the background noise of the apparatus \mbox{({\bf --})} acquired on a rigid mirror. The difference between the measurement and the background noise spectrums \mbox{({\bf \textcolor{blue}{--}})} gives a full view of the behavior of the thermal fluctuations of the cantilever in this frequency window. For comparison, the typical noise floor of a well tuned and calibrated commercial AFM is plotted \mbox{({- -})}: only the resonances could be studied, the remaining part of the spectrum would be useless.}
\label{Fig:Measurement}
\end{figure}

The measurement is performed with a home made interferometric deflection sensor \cite{Paolino-2007-instrument}, inspired by the original design of Schonenberger \cite{Schonenberger-1989} with a quadrature phase detection technique \cite{Bellon-2002-OptCom}: the interferences between the reference laser beam reflecting on the base of the cantilever and the sensing laser beam on the free end of the cantilever directly gives a calibrated measurement of the deflection $w(x=1,t)$, with very high accuracy. We illustrate in Fig.\ref{Fig:Measurement} the performance of our detection with the power spectrum density (PSD) of a rigid mirror (in black): the background noise of our system is as low as $\SI{1.3E-27}{m^{2}/Hz}$, just $\SI{50}{\%}$ higher than the shot noise limit of our detection system. Typical noise of a measurement in a commercial AFM would be at least two orders of magnitude higher.

\begin{figure}[t]
\begin{center}
\includegraphics{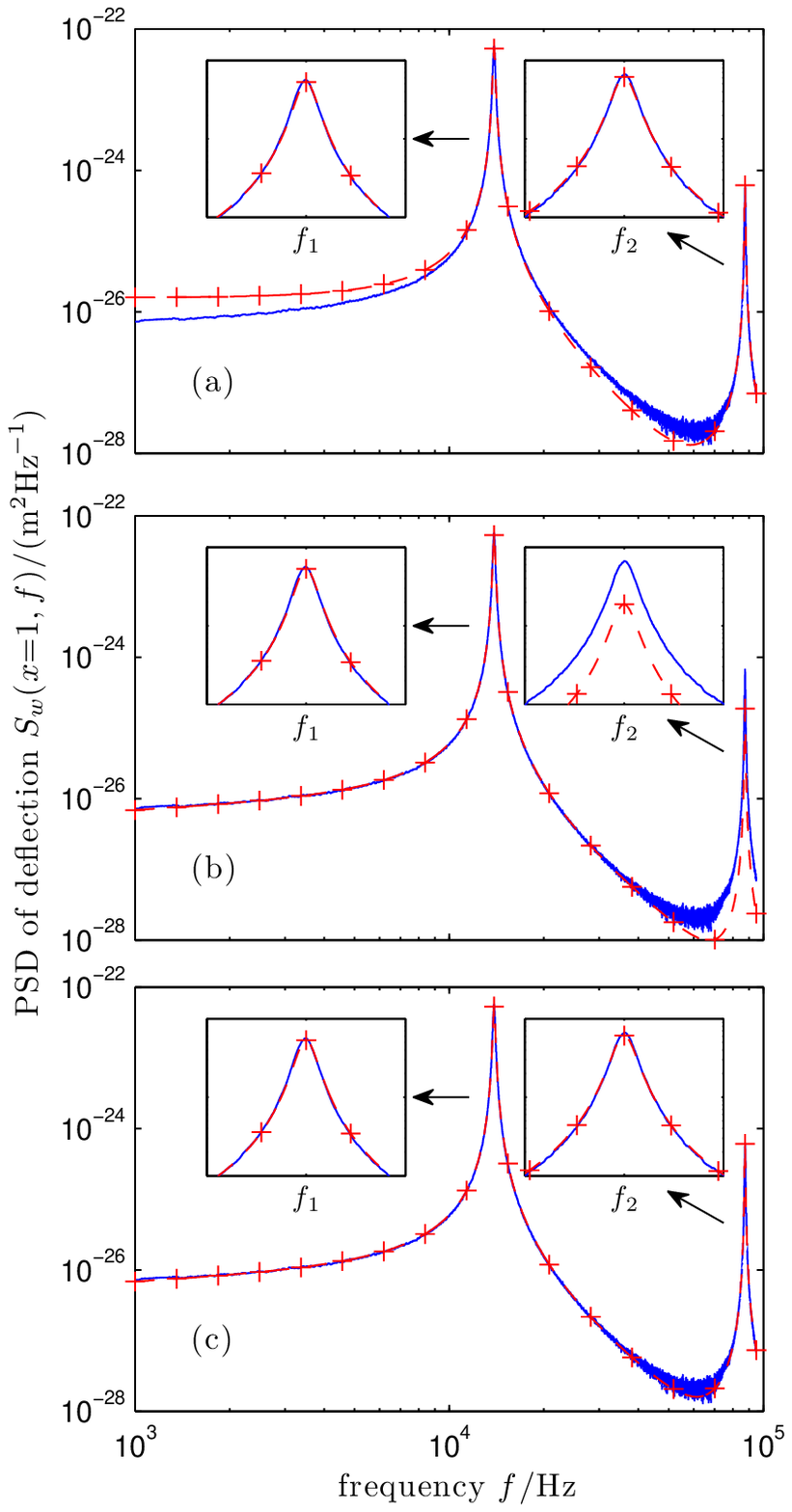}
\end{center}
\caption{Power spectrum density (PSD) of thermal noise driven deflection at the free end of the cantilever: measurement \mbox{({\bf \textcolor{blue}{---}})} vs 3 different models \mbox{({\textcolor{red}{$-\! +\! -$}})}. (a) prediction after \cite{Sader-1998}, (b) prediction after \cite{VanEysden-2007}, (c) prediction from eq.~\ref{eq:S_beta_n} and eq.~\ref{eq:PSDw} of the present work. Insets present zooms around the resonances.}
\label{Fig:Model}
\end{figure}

We plot in Fig.\ref{Fig:Measurement} the PSD of thermal noise driven deflection at the free end of the cantilever (in green). We clearly see in this spectrum the two first modes of oscillation of the mechanical beam, around $f_{1}=\SI{14}{kHz}$ and $f_{2}=\SI{87}{kHz}$, but we also have a lot of information out of these resonances as the measurement is always above the background noise of the system. If we subtract this background noise spectrum from the measurement (blue curve), we have an estimation of the thermal noise of the system on the whole $\SI{1}{kHz}$ to $\SI{100}{kHz}$ frequency interval, which we can compare to the theoretical expectation of the model presented in the previous section. The same measurement in a commercial AFM would yield much less information, as it would be limited to the resonances, and we would not have any information between them or at low frequency .

In Fig.\ref{Fig:Model}, we compare the measured PSD of fluctuations (in blue) with the prediction of 3 different models (in red)  for the thermal noise of the cantilever, all based on Sader's approach \cite{Sader-1998}. The hydrodynamic function $\Gamma(\omega)$ used for these predictions was taken from eq.~20, 21a and 21b of ref. \cite{Sader-1998}, as it correspond to the first two resonances and the extension to higher order modes of ref. \cite{VanEysden-2007} is not necessary. The first model (Fig.\ref{Fig:Model}(a)) corresponds to eq.~29a of the original article of Sader \cite{Sader-1998}, where the frequency dependence of the thermal noise forcing as been omitted. Both resonances are really well described with this model, but the off-resonance behavior is not: the low frequency noise measured is not flat as expected by the model, which also slightly underestimates the noise the intermediate region.

The second model (Fig.\ref{Fig:Model}(b)) corresponds to eq.~17a of ref. \cite{VanEysden-2007}, a corrected expression of the thermal noise proposed by Green and Sader (see ref. 44 of \cite{Green-2005}) to account for frequency dependance of the dissipation following the approach of Paul and Cross \cite{Paul-2004}. Let us first note that this expression is dimensionally incorrect, as it is homogeneous to $\SI{}{m^{2}}$ and not to $\SI{}{m^{2}/Hz}$, so we had to rescale the formula by $\SI{2.28e-4}{Hz^{-1}}$ to match the first resonance. However, this model perfectly describes the frequency dependance of the first mode, catching the slow variation of the noise in the low frequencies. The behavior of the second mode could also be well described, but it would need a higher rescaling factor.

We came to the conclusion that the scaling factor is mode dependent, and conducted the analysis of the previous part to derive the correct expression. Indeed, it can easily be shown that the frequency behavior of each mode (eq.~\ref{eq:S_beta_n}) is exactly the same as in ref.~\cite{VanEysden-2007}, but the scaling factors are different. We plot in (Fig.\ref{Fig:Model}(c)) the prediction of eq.~\ref{eq:S_beta_n} and eq.~\ref{eq:PSDw}, and see that we match the whole power spectrum density of fluctuation of deflection: both resonances, low frequency and intermediate region are very well described by the model. We stress here that the only parameter we have adjusted in the model is the thickness of the cantilever $h$, to match the actual frequencies of resonance. The Sader approach to describe the effect of the embedding fluid on the behavior of the cantilever turns out to be very good on the whole frequency range probed here.

\section{Conclusion}

In this paper, we have presented a simple theoretical framework to describe the thermal noise of a mechanical beam in a viscous fluid. Under the assumption of an infinitely thin and long cantilever, we used the Sader approach to describe the effect of the surrounding fluid (added mass and viscous drag). Using the fluctuation dissipation theorem for each flexural modes of the system, we derived a general expression for the power spectrum density of fluctuations as the sum of the contribution of the different modes. This prediction has been compared with an experimental measurement on a commercial AFM cantilever in a frequency range covering the 2 first resonances. A very good agreement has been found on the whole spectrum, with no adjustable parameters but the thickness of the cantilever. This analysis can be very useful in the AFM area, where the thermal motion of the cantilever is of great practical importance (as a lower bound to measurable forces or as a mechanical driving on its own): it is not limited to a fit of resonances only, but gives an \emph{a priori} knowledge of the cantilever noise on the whole frequency spectrum.

\bigskip

{\bf Acknowledgements}

I thank F. Vittoz and F. Ropars for technical support, and P. Paolino, S. Ciliberto, A. Petrosyan, J.P. Aim\'e and F. Bertin for stimulating discussions. This work has been partially supported by contract ANR-05-BLAN-0105-01 of the Agence Nationale de la Recherche in France.

\end{document}